\documentclass{PoS}
\usepackage{graphicx}
\usepackage{amssymb,amsmath}
\usepackage{slashed}
\usepackage{url}

\newcommand{\plotscale}{.49}

\title{Neutrino scattering with nuclei --- Theory of low energy nuclear effects and its applications}

\ShortTitle{Neutrino scattering with nuclei}

 \author{Tina Leitner\thanks{Speaker in WG2: Nuclear Effects in Neutrino Scattering I.}
   \\
          Universit\"at Giessen, Germany\\
         E-mail: \email{tina.j.leitner@theo.physik.uni-giessen.de}}

 \author{Oliver Buss  \\
          Universit\"at Giessen, Germany\\
         E-mail: \email{oliver.buss@theo.physik.uni-giessen.de}}

 \author{Ulrich Mosel\thanks{Speaker in plenary session.}  \\
          Universit\"at Giessen, Germany\\
         E-mail: \email{ulrich.mosel@theo.physik.uni-giessen.de}}

 \author{Luis Alvarez-Ruso\\
         Universidad de Murcia, Spain\\
       E-mail: \email{luis.alvarez@ific.uv.es}}

\abstract{Current long baseline (LBL) experiments aim at measuring neutrino oscillation parameters with a high precision.
A critical quantity is the neutrino energy which can not be measured directly but has to be reconstructed from the observed hadrons. A good knowledge of neutrino-nucleus interactions is thus necessary to minimize the systematic uncertainties in neutrino fluxes, backgrounds and detector responses.
In particular final-state interactions inside the target nucleus modify considerably the particle yields through rescattering, charge-exchange and absorption.
Nuclear effects can be described with our coupled channel GiBUU transport model where the neutrino first interacts with a bound nucleon producing secondary particles which are then transported out of the nucleus. In this contribution, we give some examples for the application of our model focusing in particular on the MiniBooNE and K2K experiments.}

\FullConference{10th International Workshop on Neutrino Factories, Super beams and Beta beams\\
		 June 30 - July 5 2008\\
		 Valencia, Spain}

\begin{document}


\newcommand{\myunit}[1]{\mbox{$\,\text{#1}$}}
\newcommand{\MeV}{\myunit{MeV}}
\newcommand{\GeV}{\myunit{GeV}}
\newcommand{\GeVsq}{\mbox{$\,\text{GeV}^2$}}
\newcommand{\GeVminsq}{\mbox{$\,\text{GeV}^{-2}$}}
\newcommand{\cmsq}{\mbox{$\,\text{cm}^2$}}
\newcommand{\nucmsq}{\mbox{$\,10^{-38}$\cmsq}}

\newcommand{\myvec}[1]{\mathbf{#1}}

\newcommand{\reffig}[1]{Fig.~\ref{#1}}
\newcommand{\refeq}[1]{Eq.~(\ref{#1})}
\newcommand{\refch}[1]{Chapter~\ref{#1}}
\newcommand{\refsec}[1]{Section~\ref{#1}}
\newcommand{\refapp}[1]{Appendix~\ref{#1}}
\newcommand{\refcite}[1]{Ref.~\cite{#1}}
\newcommand{\refscite}[1]{Refs.~\cite{#1}}
\newcommand{\refetal}[1]{\emph{et~al.}~\cite{#1}}
\newcommand{\reftab}[1]{Table~\ref{#1}}

\newcommand{\kpi}{k_\pi}
\newcommand{\kpiz}{k_\pi^0}
\newcommand{\slashkpi}{\slashed{k}_\pi}
\newcommand{\kpicn}[1]{k_\pi^{#1}}
\newcommand{\kpico}[1]{{k_\pi}_{#1}}
\newcommand{\abskpi}{|\myvec{\kpi}|}

\newcommand{\kz}{k^0}
\newcommand{\kcn}[1]{k^{#1}}
\newcommand{\kco}[1]{k_{#1}}
\newcommand{\slashk}{\slashed{k}} 
\newcommand{\absk}{|\myvec{k}|}
\newcommand{\enu}{E_\nu}

\newcommand{\kpr}{k'}
\newcommand{\kprz}{{k'}^0}
\newcommand{\kprcn}[1]{k'\,^{#1}}
\newcommand{\kprco}[1]{k'_{#1}}
\newcommand{\slashkpr}{\slashed{k}'} 
\newcommand{\abskpr}{|\myvec{\kpr}|}
\newcommand{\sinth}{\sin \theta}
\newcommand{\costh}{\cos \theta}
\newcommand{\sinph}{\sin \phi}
\newcommand{\cosph}{\cos \phi}
\newcommand{\elep}{E_\lep}

\newcommand{\pz}{p^0}
\newcommand{\pcn}[1]{p^{#1}}
\newcommand{\pco}[1]{p_{#1}}
\newcommand{\slashp}{\slashed{p}} 
\newcommand{\absp}{|\myvec{p}|}

\newcommand{\qz}{q^0}
\newcommand{\slashq}{\slashed{q}} 
\newcommand{\absq}{|\myvec{q}|}

\newcommand{\ppr}{p'}
\newcommand{\pprz}{{p'}^0}
\newcommand{\pprcn}[1]{p'\,^{#1}}
\newcommand{\pprco}[1]{p'_{#1}}
\newcommand{\slashppr}{\slashed{p}'} 
\newcommand{\absppr}{|\myvec{\ppr}|}

\newcommand{\M}{M}
\newcommand{\Mpr}{M'\hspace{.2mm}}
\newcommand{\mpi}{m_\pi}
\newcommand{\mnu}{m_\nu}
\newcommand{\ml}{m_\ell}
\newcommand{\mlpr}{m_{\ell'}}

\newcommand{\tinyMM}[1]{\text{\begin{tiny}#1\end{tiny}}}
\newcommand{\ME}{\mathcal{M}}
\newcommand{\Res}{R}

\newcommand{\subQE}{\text{QE}} 
\newcommand{\subhalf}{\text{\begin{tiny}1/2\end{tiny}}}
\newcommand{\subthreehalf}{\text{\begin{tiny}3/2\end{tiny}}}
\newcommand{\subCC}{\text{CC}}
\newcommand{\subNC}{\text{NC}}
\newcommand{\subEM}{\text{EM}}
\newcommand{\subBG}{\text{BG}}
\newcommand{\subtot}{\text{tot}}

\newcommand{\g}[1]{\gamma^{#1}}
\newcommand{\gF}{\gamma^5}
\newcommand{\gN}{\gamma^0}
\newcommand{\gmu}{\gamma^\mu}

\newcommand{\Tr}{\mbox{Tr}}
\newcommand{\pr}[1]{{#1}'}
\newcommand{\RealPart}{\text{Re}}
\newcommand{\ImaginaryPart}{\text{Im}}

\newcommand{\ii}{\mathrm{i}}
\newcommand{\dd}{\mathrm{d}}
\newcommand{\dfp}[1]{ \frac{\dd^4 {#1} }{(2\pi)^3} }
\newcommand{\dsdqs}{\frac{\dd \sigma}{\dd Q^2}}
\newcommand{\fff}{\mathcal{F}}
\newcommand{\ffc}{\mathcal{C}}
\newcommand{\jj}{\mathcal{J}}

\newcommand{\atom}[2]{\mbox{$^{#1}\text{#2}$}}
\newcommand{\carbon}{{\atom{12}{C}}}
\newcommand{\oxygen}{{\atom{16}{O}}}
\newcommand{\calcium}{{\atom{40}{Ca}}}
\newcommand{\iron}{{\atom{56}{Fe}}}

\newcommand{\res}[3]{#1$_{#2}$(#3)}
\newcommand{\lep}{\ell}
\newcommand{\coscab}{\mbox{$\cos \theta_C$}}
\newcommand{\sinwein}{\mbox{$\sin^2 \theta_W$}}

\newcommand{\matrixelement}[3]{\left \langle {#1} \left\lvert \vphantom{ {#1} {#2} {#3}} {#2} \right\rvert {#3} \right\rangle}
\newcommand{\clebsch}[2]{\left( {#1} \left\lvert \vphantom{{#1}{#2}} \right. {#2} \right)}

\section{Introduction}

There is an extensive experimental effort aiming at a precise determination of neutrino oscillation parameters. However, neutrino oscillation results depend on the neutrino energy---a quantity which can not be measured directly but has to be reconstructed from the hadronic debris coming out of the neutrino-nucleus reaction inside the detector. A reliable reconstruction of the neutrino kinematics and the initial scattering process has to account for in-medium modifications and, in particular, for final state interactions inside the target nucleus. They can, e.g., through intra-nuclear rescattering, change particle multiplicities and also redistribute their energy.

Those effects can be simulated with our fully coupled channel GiBUU transport model where the neutrino first interacts with a bound nucleon producing secondary particles which are then transported out of the nucleus. We use a formalism that incorporates recent form factor parametrizations and apply, besides Fermi motion and Pauli blocking, important ingredients of the many-body problem such as mean-field potentials, in-medium spectral functions and RPA correlations. The modeling of final state interactions includes a large variety of possible interactions channels and, furthermore, particles with an in-medium width are transported off-shell.

This article is structured in the following way: First we introduce our model for the interaction of neutrinos and electrons with bound nucleons. We then outline our transport model used to describe final state interactions (FSI). Thereafter, some sample results for $\nu A$ scattering are presented. Finally, we apply our model to describe recent observations of the LBL experiments MiniBooNE and K2K.

\section{GiBUU model}

Lepton induced scattering in the GiBUU model is treated as a two step process: First, the leptons scatter of nucleons embedded in the nuclear medium. Then, the outcome of this initial reaction is propagated through the nucleus, using a hadronic transport approach. More details can be found in \refcite{gibuu}.

\subsection{Initial vertex}

We focus on the charged current (CC) ($\nu N \rightarrow \ell^- X$) reaction, but discuss also the electromagnetic (EM) ($\ell^- N \rightarrow \ell^- X $) one used as a benchmark for our neutrino calculations.

We treat the nucleus as a local Fermi gas of nucleons bound in a mean field potential. The total reaction rate for the scattering of a lepton with four-momentum $k=(k_0,\myvec{k})$ off a nucleon with momentum $p=(E,\myvec{p})$, going into a lepton with momentum $k'=(k'_0,\myvec{k'})$ is given by an incoherent sum over all nucleons (impulse approximation)
\begin{equation}
\frac{\dd\sigma_\tinyMM{EM,CC}}{\dd \omega \; \dd\Omega}
=\sum^A_{j=1} \left( \frac{\dd\sigma^\tinyMM{tot}_\tinyMM{EM,CC}}{\dd \omega \; \dd\Omega} \right)_{\hspace{-1.6mm}j}, \label{eq:xsecTotal}
\end{equation}
with $\omega=k_0-k'_0$, $Q^2=-(k-k')^2$ and $\Omega = \angle (\myvec{k},\myvec{k'})$. The cross sections on the rhs of \refeq{eq:xsecTotal} are medium-modified (see below).

\paragraph{Elementary input cross sections.}
In the intermediate energy region ($k_0 \sim 0.5-2 \GeV$), the cross section is dominated by quasielastic (QE) scattering ($ e  N \to e'  N'$ and $\nu  N \to \ell^-  N'$) and resonance excitation\footnote{mainly \res{P}{33}{1232}} ($ e  N \to e'  R$ and $\nu  N \to \ell^-  R$). Furthermore, we account for non-resonant single-pion backgrounds for both $ e  N \to e'  \pi  N'$ and $\nu  N \to \ell^-  \pi  N'$. Thus we assume
\begin{equation}
\frac{\dd\sigma^\tinyMM{tot}_\tinyMM{EM,CC}}{\dd \omega \; \dd\Omega}=\frac{\dd\sigma^\tinyMM{QE}_\tinyMM{EM,CC}}{\dd \omega \; \dd\Omega}+\sum_R \frac{\dd\sigma^R_\tinyMM{EM,CC}}{\dd \omega \; \dd\Omega} + \frac{\dd\sigma^\tinyMM{BG}_\tinyMM{EM,CC}}{\dd \omega \; \dd\Omega}, \label{eq:xsecTotalContri}
\end{equation}
where $\dd \sigma^\tinyMM{BG}_\tinyMM{EM,CC}$ also contains contributions from resonance-background interference.

Omitting phase space factors, the cross section for QE scattering and resonance excitation is given by \cite{Buss:2007ar}
\begin{equation}
\frac{d\sigma^\tinyMM{QE,R}_\tinyMM{EM,CC}}{d \omega \; d \Omega} \propto  \mathcal{A}(E',\myvec{p'}) \;  L_{\mu \nu} H^{\mu \nu}_\tinyMM{QE,R}, \label{eq:QEres_cross}
\end{equation}
where $p'=(E',\myvec{p'})$ is the four-momentum of the outgoing nucleon and  $\mathcal{A}(E',\myvec{p'})$ gives the spectral function for the outgoing baryon. $L_{\mu \nu}$ is the leptonic tensor.

The QE hadronic tensor $H^{\mu \nu}_\tinyMM{QE}$ can be parametrized in terms of vector and axial form factors (see, e.g., our earlier work \cite{Leitner:2006ww}). The vector form factors are taken from the latest analysis by Bodek \refetal{Bodek:2007vi}; a dipole ansatz with $M_A=0.999\GeV$ \cite{Kuzmin:2007kr} is used for the axial ones.

The resonance hadronic tensor $H^{\mu \nu}_\tinyMM{R}$ depends on the specific resonance.
For spin 1/2 resonances with positive parity (e.g.~\res{P}{11}{1440}) we find for the hadronic current $J^{\mu}_{1/2+}=V^{\mu}_{1/2} - A^{\mu}_{1/2}$ with
\begin{equation}
V^{\mu}_{1/2}-A^{\mu}_{1/2}=\frac{F_1^V}{(2 M_N)^2} \left( Q^2 \gamma^\mu + \slashq q^\mu \right) + \frac{F_2^V}{2 M_N} i \sigma^{\mu\alpha} q_\alpha + F_A \gamma^\mu \gamma_5 + \frac{F_P}{M_N} q^\mu \gF,
\end{equation}
and for states with negative parity (e.g.~\res{S}{11}{1535}) we use $J^{\mu}_{1/2-}=[ V^{\mu}_{1/2} - A^{\mu}_{1/2} ] \gF$.

For spin 3/2 resonances with positive parity as the \res{P}{33}{1232}, we have $J^{\alpha \mu }_{3/2+} = [ V^{\alpha \mu }_{3/2} - A^{\alpha \mu }_{3/2}] \gamma_{5}$ with
\begin{equation}
V^{\alpha \mu }_{3/2} =
  \frac{C_3^V}{M_N} (g^{\alpha \mu} \slashq - q^{\alpha} \gamma^{\mu})+
  \frac{C_4^V}{M_N^2} (g^{\alpha \mu} q\cdot \ppr - q^{\alpha} {\ppr}^{\mu}) + \frac{C_5^V}{M_N^2} (g^{\alpha \mu} q\cdot p - q^{\alpha} p^{\mu}) + g^{\alpha \mu} C_6^V \label{eq:vectorspinthreehalfcurrent}
\end{equation}
and
\begin{equation}
-A^{\alpha \mu }_{3/2} = \left[\frac{C_3^A}{M_N} (g^{\alpha \mu} \slashq - q^{\alpha} \gamma^{\mu})+
  \frac{C_4^A}{M_N^2} (g^{\alpha \mu} q\cdot \ppr - q^{\alpha} {\ppr}^{\mu})+
 {C_5^A} g^{\alpha \mu}  + \frac{C_6^A}{M_N^2} q^{\alpha} q^{\mu}\right] \gamma_{5};\label{eq:axialspinthreehalfcurrent}
\end{equation}
for the ones with negative parity (e.g.~\res{D}{13}{1535}) we use $J^{\alpha \mu }_{3/2-} = V^{\alpha \mu }_{3/2} - A^{\alpha \mu }_{3/2}$.
As an approximation, resonances with spin greater than 3/2 are treated within the spin 3/2 formalism.

The vector form factors $F_{1,2}^V$ ($C_{3,4,5,6}^V$) present in CC scattering are related to the electromagnetic transition form factors $F^N_{1,2}$ ($C_{3,4,5,6}^N$) with $N=p,n$ and those again to helicity amplitudes~\cite{Fogli:1979cz,Alvarez-Ruso:1997jr,Lalakulich:2006sw,Hernandez:2007ej,inProgress}, which can be extracted from electron scattering experiments. We apply the results of the recent MAID2005 analysis \cite{Tiator:2006dq} which includes 13 resonances with $W < 2$ GeV; all of them are implemented in our model---a list is given in \reftab{tab:included_resonances}.
\begin{table}
\begin{center}
\begin{tabular}{c c c c c c c c}
\hline \hline
name               & $M_R$ [GeV] & J   & I   & P & $\Gamma_0^{\text{tot}}$ [GeV]&  $\Gamma_0^{\pi N}/\Gamma_0^{\text{tot}}$   & axial coupling \\ \hline
\res{P}{33}{1232}  &       1.232 & 3/2 & 3/2 & + &                 0.118 &                               1.00 & 1.17\\
\res{P}{11}{1440}  &       1.462 & 1/2 & 1/2 & + &                 0.391 &                               0.69 & -0.52\\
\res{D}{13}{1520}  &       1.524 & 3/2 & 1/2 & - &                 0.124 &                               0.59 &-2.15\\
\res{S}{11}{1535}  &       1.534 & 1/2 & 1/2 & - &                 0.151 &                               0.51 &-0.23\\
\res{S}{31}{1620}  &       1.672 & 1/2 & 3/2 & - &                 0.154 &                               0.09 &-0.05\\
\res{S}{11}{1650}  &       1.659 & 1/2 & 1/2 & - &                 0.173 &                               0.89 &-0.25\\
\res{D}{15}{1675}  &       1.676 & 5/2 & 1/2 & - &                 0.159 &                               0.47 &-1.38\\
\res{F}{15}{1680}  &       1.684 & 5/2 & 1/2 & + &                 0.139 &                               0.70 & 0.43\\
\res{D}{33}{1700}  &       1.762 & 3/2 & 3/2 & - &                 0.599 &                               0.14 & -0.84\\
\res{P}{13}{1720}  &       1.717 & 3/2 & 1/2 & + &                 0.383 &                               0.13 & 0.29\\
\res{F}{35}{1905}  &       1.881 & 5/2 & 3/2 & + &                 0.327 &                               0.12 & 0.15\\
\res{P}{31}{1910}  &       1.882 & 1/2 & 3/2 & + &                 0.239 &                               0.23 &-0.08\\
\res{F}{37}{1950}  &       1.945 & 7/2 & 3/2 & + &                 0.300 &                               0.38 & 0.24\\
\hline \hline
\end{tabular}
\end{center}
\caption{Properties of the resonances included in our model. The pole mass $M_R$, spin $J$, isospin $I$, parity $P$, the vacuum total decay width $\Gamma_0^{\text{tot}}$, the branching ratio into $\pi N$ and the axial coupling are listed (see text for details on the extraction of the axial coupling). The resonance parameters are taken from \refcite{manley}. \label{tab:included_resonances}}
\end{table}

The lack of precise data renders the determination of the axial form factors difficult. Pion pole dominance and the PCAC hypothesis allow on one side to relate $F_P$ to $F_A$ ($C_6^A$ to $C_5^A$) and on the other side to extract the axial coupling $F_A(0)$ ($C_5^A(0)$) \cite{inProgress} (given in \reftab{tab:included_resonances}). We assume a dipole form with $M_A^*=1\GeV$ for $F_A$ and all $C_5^A$ except for the \res{P}{33}{1232}. $C_3^A$ and $C_4^A$ are set to zero for all resonances except the \res{P}{33}{1232}.

\begin{figure}[tbp]
  \centerline{\includegraphics[scale=\plotscale]{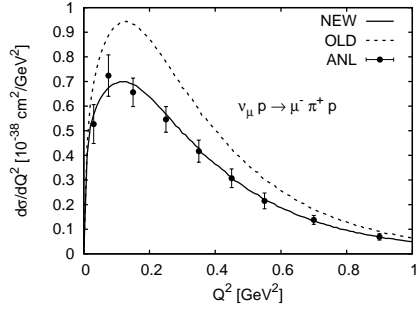}
  }
  \caption{Differential cross section $\dsdqs$ averaged over the ANL flux for two different sets of parameters describing $C_5^A$ compared to the ANL data \cite{Radecky:1981fn}. An invariant mass cut $W<1.4\GeV$ is applied. \label{fig:P33_1232_axFF_fit_to_Qs}}
\end{figure}

For the $\Delta$ resonance, some experimental information is available from ANL~\cite{Barish:1978pj,Radecky:1981fn} and BNL~\cite{Kitagaki:1986ct}. Applying the Adler model~\cite{Adler:1968tw} where $C_4^{A}(Q^2)=-{C_5^{A}(Q^2)}/{4}$ and $C_3^{A}(Q^2)= 0$ we can extract the $Q^2$ dependence of $C_5^A$ from these data. Any update of the vector form factors requires the axial ones to be refitted. Improving on the electromagnetic vector form factors without readjusting the axial ones \cite{Lalakulich:2006sw} will result in a worse description of the neutrino data.
Assuming that PCAC holds (i.e.~the value of $C_5^{A}(0)$ is unchanged\footnote{This is in contrast to the work of Hernandez \refetal{Hernandez:2007qq} which takes it as a free parameter. Since we rely on PCAC for all other resonance excitations (where no data are available) we prefer to keep it here, too. Furthermore, this coupling was extracted from the BNL data in \refcite{Alvarez-Ruso:1998hi} and found to be consistent with the PCAC prediction.}) and neglecting the small non-resonant background in the channel $\nu p \to \mu^- \pi^+ p$, we find
\begin{equation}
C_5^A(Q^2) = {{C_5^A(0)\left[ 1+\frac{a Q^2}{b+Q^2} \right] } {\left( 1+ \frac{Q^2}{{M_A^\Delta}^2}\right)^{-2}}} ,
\end{equation}
with $a=-0.25$ and $b=0.04\GeVsq$ and $M_A^\Delta=0.95\GeV$ (set ``NEW''). Formerly, $a=-1.21$ and $b=2\GeVsq$ and $M_A^\Delta=1.28\GeV$ (set ``OLD'') were used \cite{Barish:1978pj,Radecky:1981fn,Kitagaki:1986ct}.
With the new parameters, good agreement with the ANL data is reached (solid line in \reffig{fig:P33_1232_axFF_fit_to_Qs}).
In order to illustrate the sensitivity of the cross section to the refitted axial form factor, we show in \reffig{fig:P33_1232_axFF_fit_to_Qs} also the dashed curve which gives the cross section obtained with the new vector form factors but the old axial ones.

\begin{figure}
         \centerline{\includegraphics[scale=\plotscale]{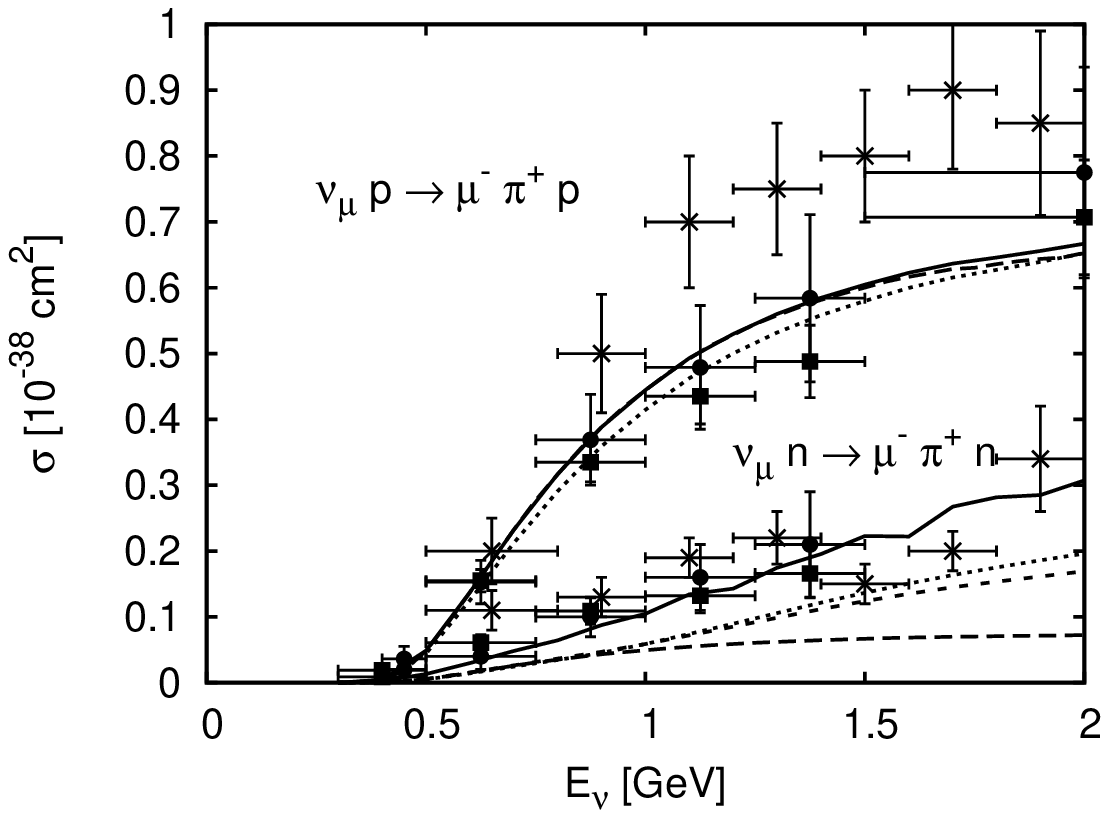}
 	\hspace{1em}
         \includegraphics[scale=\plotscale]{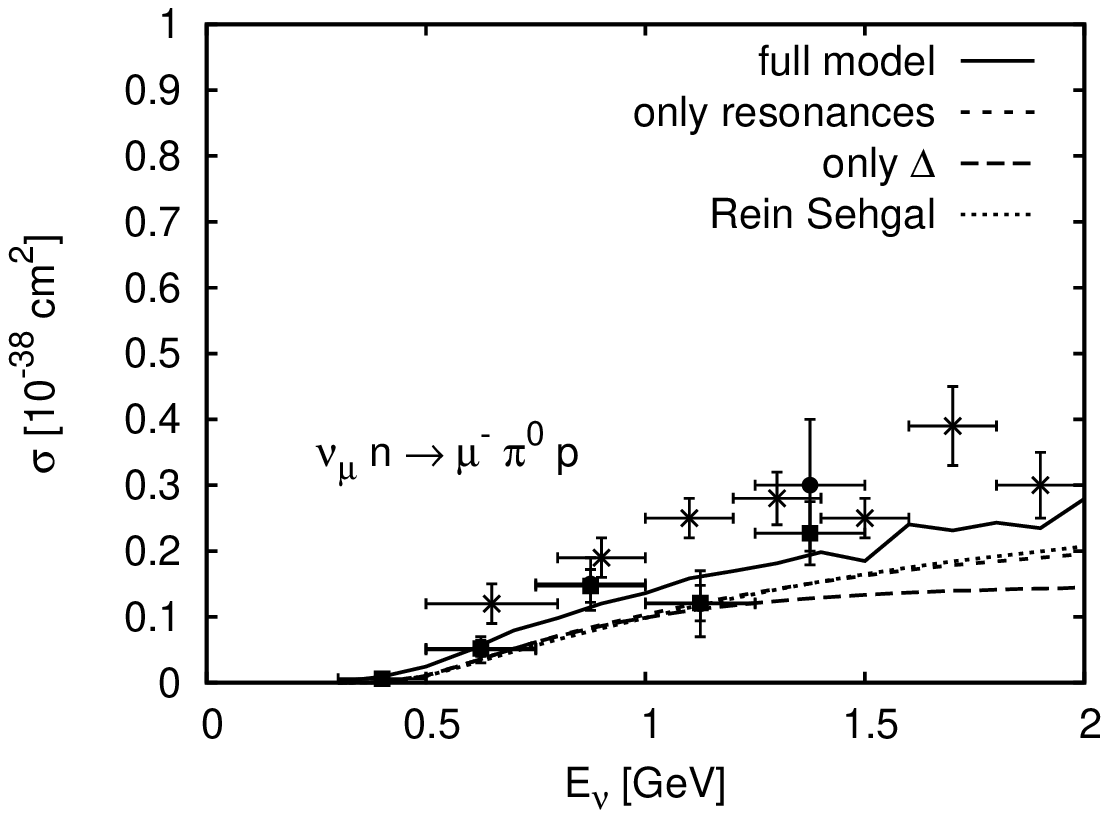}}
       \caption{Total $\pi^+$ (left) and $\pi^0$ (right) production cross sections through resonance excitation and non-resonant processes as functions of the neutrino energy (solid lines) compared to the pion production data of of ANL (Refs.~\cite{Barish:1978pj} ($\bullet$) and \cite{Radecky:1981fn} ($\blacksquare$)) and BNL (\cite{Kitagaki:1986ct} ($\times$)). The different contributions are indicated by \emph{only resonances} (short-dashed) and \emph{only $\Delta$} (long-dashed). The result obtained in the model of Rein and Sehgal is also shown (dotted).  \label{fig:xsec_CCpionprod}}
 \end{figure}

In the case of electro-production, there is a wealth of data which allows to determine the single pion background $\dd\sigma^\tinyMM{bg}_\tinyMM{EM}/\dd \omega \dd\Omega$. This is done by subtracting the dominant resonance contribution from the total single-pion cross section \cite{Buss:2007ar,Buss:2008:Bormio}
\begin{equation}
\frac{\dd\sigma^\tinyMM{bg}_\tinyMM{EM}}{\dd \omega \; \dd\Omega}
=\frac{\dd\sigma^{1\pi}_\tinyMM{EM}}{\dd \omega \; \dd\Omega}-\sum_R\frac{\dd\sigma^{R}_\tinyMM{EM}}{\dd \omega \; \dd\Omega}.
\label{eq:bgDef}
\end{equation}
The total single-pion production cross section on the nucleon $\dd\sigma^{1\pi}_\tinyMM{EM}/\dd \omega  \dd\Omega$ is taken from MAID~\cite{Tiator:2006dq}.
Such a treatment is not possible in the neutrino case, since there are not enough experimental data to fix the additionally necessary six axial amplitudes, hence some simplifications are required\footnote{Besides the phenomenological ansatz we are following in this work, one can apply elementary models to estimate the single-pion non-resonant terms~\cite{Fogli:1979cz,Hernandez:2007qq,Sato:2003rq}.} and we assume
\begin{equation}
\frac{\dd\sigma^\tinyMM{bg}_\tinyMM{CC}}{\dd \omega \; \dd\Omega} =\left( 1 +b^{N \pi}\right) \frac{\dd\sigma^{V}_\tinyMM{CC}}{\dd \omega \; \dd\Omega},
\end{equation}
where $\dd\sigma^{V}_\tinyMM{CC}$ is constraint by electron scattering data. The factor $b^{N \pi}$ depends on the channel, $\nu n \to l^- n \pi^+$ or $\nu n \to l^- p \pi^0$ ($\nu p \to \mu^- \pi^+ p$ is assumed to be ``background-free''): with $b^{p \pi^0}=3$ and $b^{n \pi^+}=1.5$ a reasonable agreement with the ANL data is reached as can be seen from \reffig{fig:xsec_CCpionprod}.

\begin{figure}
         \centerline{\includegraphics[scale=\plotscale]{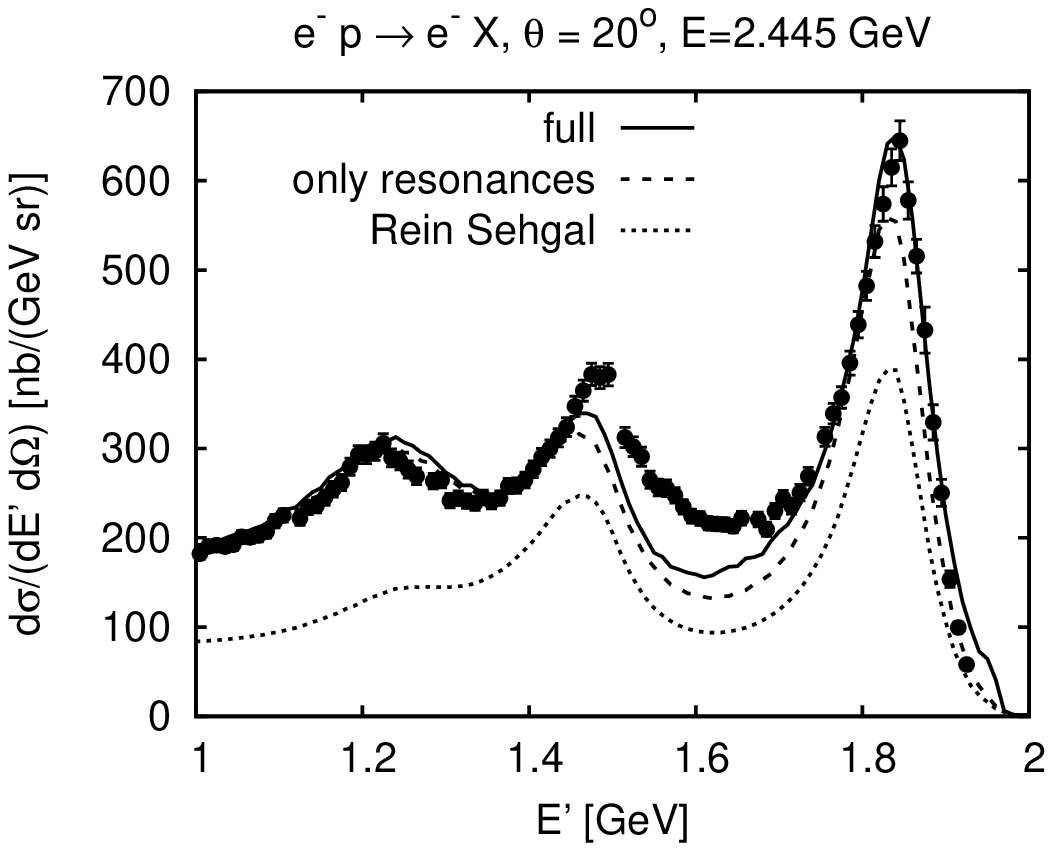}
 	\hspace{1em}
         \includegraphics[scale=\plotscale]{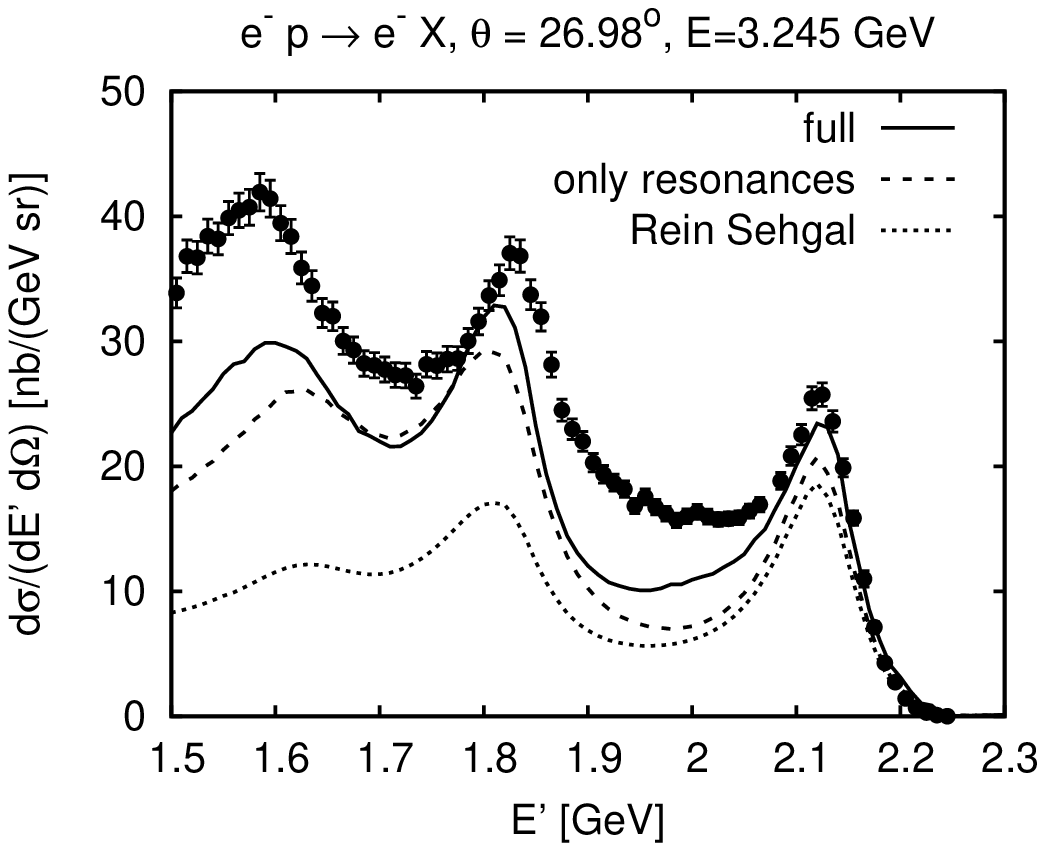}}
       \caption{Double differential cross sections for scattering of electrons off protons through resonance excitation and non-resonant processes as a function of the energy of the outgoing electron (solid lines). The dashed lines show the resonance contribution, and the dotted lines the outcome of the model of Rein and Sehgal. We compare to data from JLAB \cite{JLABWebsite}.  \label{fig:xsec_EMinclproton}}
 \end{figure}

The full cross section \refeq{eq:xsecTotalContri} is shown in \reffig{fig:xsec_EMinclproton} for electron scattering off protons. The right peak is dominated by the $\Delta$, the second and third resonance region are clearly visible (middle and left peak), the QE peak is not shown. The different contributions to the cross section are shown and compared to data. The inclusion of the $1\pi$ non-resonant background is necessary, but not sufficient to achieve a good description of the data at higher bombarding energies (right panel); multi-pion backgrounds should also be considered in the future. We further compare to the model of Rein and Sehgal \cite{Rein:1980wg}, a model widely used in neutrino event generators, and find that this model underestimates significantly the electron data as observed also by Graczyk \refetal{Graczyk:2007bc}.
In \reffig{fig:xsec_CCpionprod} we plot the neutrino induced total $\pi^+$ (left panel) and $\pi^0$ (right panel) production cross sections; the different contributions are also indicated (full: resonances + background). The excitation of higher resonances is almost invisible in the isospin 3/2 channel (upper curves in the left panel, solid vs.~long-dashed line). Again, we compare to Rein and Sehgal which gives for these integrated cross sections results similar to ours (comparing the dotted line to our resonance contribution).

\paragraph{Medium modifications.}

The target nucleus is treated within a local Thomas-Fermi approximation as a Fermi gas of nucleons bound by a mean-field potential $U_{S_N}(\myvec{p},\myvec{r})$ which is parametrized as a sum of a Skyrme term depending only on density and a momentum-dependent contribution.

The spectral function of a particle with four-momentum $p=(E,\myvec{p})$ and mass $M=\sqrt{p^2}$ is given by
\begin{equation}
 \mathcal{A}(E,\myvec{p})=\frac{1}{\pi}\frac{-\ImaginaryPart \Sigma(E,\myvec{p})}{(M^2-M_0^2-\RealPart \Sigma(E,\myvec{p}))^2+(\ImaginaryPart \Sigma(E,\myvec{p}))^2},
\end{equation}
with the self energy $\Sigma(E,\myvec{p})$ and the vacuum pole-mass $M_0$.
It includes the effect of the momentum-dependent potential on the outgoing baryons and also accounts for the in-medium collisional broadening of the outgoing final states. We neglect the spectral functions of the initial states because their widths are considerably smaller than those of the outgoing nucleons \cite{CiofidegliAtti:1990vn}.
The imaginary part of the self energy is related to the full width, $\Gamma_\text{tot}$, in the medium, via $\ImaginaryPart \Sigma(E,\myvec{p})=-M \Gamma_\text{tot},$ which is given by $\Gamma_\text{tot}=\Gamma_\text{PB}+\Gamma_\text{coll}$.
Due to Pauli blocking (PB) of the final state particles in the medium, the free decay width is lowered. On the other side, both the nucleons and the $\Delta$ resonances undergo collisions with the nucleons in the Fermi sea. This leads to a collisional broadening of the particle width. To estimate this collisional broadening, we apply the low-density approximation
\begin{equation}
\Gamma_\text{coll}(E,\myvec{p})
=\sum_{n,p}\; \int_{\text{FS}} \\ \left. \sigma(E,\myvec{p},\myvec{p'})~v_\text{rel}~P_\text{PB} \right.~\frac{\dd^3p'}{(2 \pi)^3} , \label{eq:specfunc}
\end{equation}
where we integrate over all nucleon momenta in the Fermi sphere (FS). $\sigma(E,\myvec{p},\myvec{p'})$ denotes the total cross section for the scattering of the outgoing nucleon/resonance with a nucleon of momentum $\myvec{p'}$ in the vacuum; $v_\text{rel}$ denotes the relative velocity of the particle and the nucleon, $P_\text{PB}$ is the Pauli blocking factor for the final state particles. The total cross sections are chosen according to the GiBUU collision term (for details see \cite{gibuu}).
The real part of the self-energy is given as once-subtracted dispersion relation where the subtraction point is fixed by the mean fields. This procedure guarantees the proper normalization of the spectral functions \cite{Buss:2007ar}.

In the nucleus, the elementary cross sections discussed above for resonance excitation and quasielastic scattering are evaluated with full in-medium kinematics accounting for the momen\-tum-dependent mean field. Furthermore, also the flux and phase-space factors are evaluated with in-medium four-vectors.
As an approximation, we use in the medium the same form-factor para\-metri\-zations as in vacuum.
Pauli blocking is taken into account by multiplying each cross section on the rhs of \refeq{eq:xsecTotal} with the Pauli-blocking factor.
In particular, the momentum dependence of the potential and the collisional broadening of the nucleon improve the correspondence with the data considerably~\cite{Buss:2007ar}.

\subsection{Final state interactions}

The final-state interactions (FSI) of the produced particles are implemented by means of the coupled-channel semi-classical Giessen Boltzmann-Uehling-Uhlenbeck (GiBUU) transport model \cite{gibuu}. Originally developed to describe heavy-ion collisions~\cite{Teis:1996kx}, it has been extended to describe the interactions of pions, real and virtual photons and neutrinos with nuclei~\cite{Buss:2007ar,Leitner:2006ww,Buss:2008:Bormio,Effenberger:1999ay, Falter:2004uc, Leitner:2006sp, Buss:2006vh, Buss:2006yk}.

In this model, we describe the space-time evolution of a many-particle system under the influence of a mean-field potential and a collision term by a BUU equation for each particle species. A collision term accounts for changes (gain and loss) in the phase space density due to elastic and inelastic collisions between the particles, and also due to particle decays into other hadrons whenever it is allowed by Pauli blocking. The most relevant states for neutrino-induced reactions at intermediate energies are the nucleon, the $\Delta$~resonance and the pion. For the $NN$ cross section and its angular dependence we use a fit to data from \refcite{Cugnon:1996kh}. For the pion cross sections we use a resonance model with the background fitted to data as shown in detail in \refcite{Buss:2006vh}. The decay of resonances into a pion nucleon pair is Pauli blocked if the momentum of the nucleon is below the Fermi momentum. We allow not only for the decay of the resonances, but also for the rescattering in the nuclear medium through processes like $R N \to N N$, $R N \to R' N$ and, for the $\Delta$ resonance we also consider $\Delta N N \to N N N$ based on \cite{Oset:1987re}.

In between the collisions, all particles (also resonances) are propagated in their mean-field potential according to their BUU equation. We emphasize again, that, due to rescattering effects in the medium, the nucleon and the resonances acquire an additional complex self energy leading to modified spectral functions obtained in a consistent way from the GiBUU cross sections (see \refeq{eq:specfunc}). Thus, the nucleon and the resonances are transported off-shell in our model. Thereby we ensure that the particles are transported back to their vacuum spectral function when leaving the nucleus.

In conclusion, FSI lead to absorption, charge exchange and redistribution of energy and momentum as well as to the production of new particles. In our coupled-channel treatment of the FSI --- in which the BUU equations are coupled through the collision term and, with less strength, also through the potentials---our model differs from standard Glauber approaches that do not allow for side-feeding and rescattering.

\begin{figure}
         \centerline{\includegraphics[scale=\plotscale]{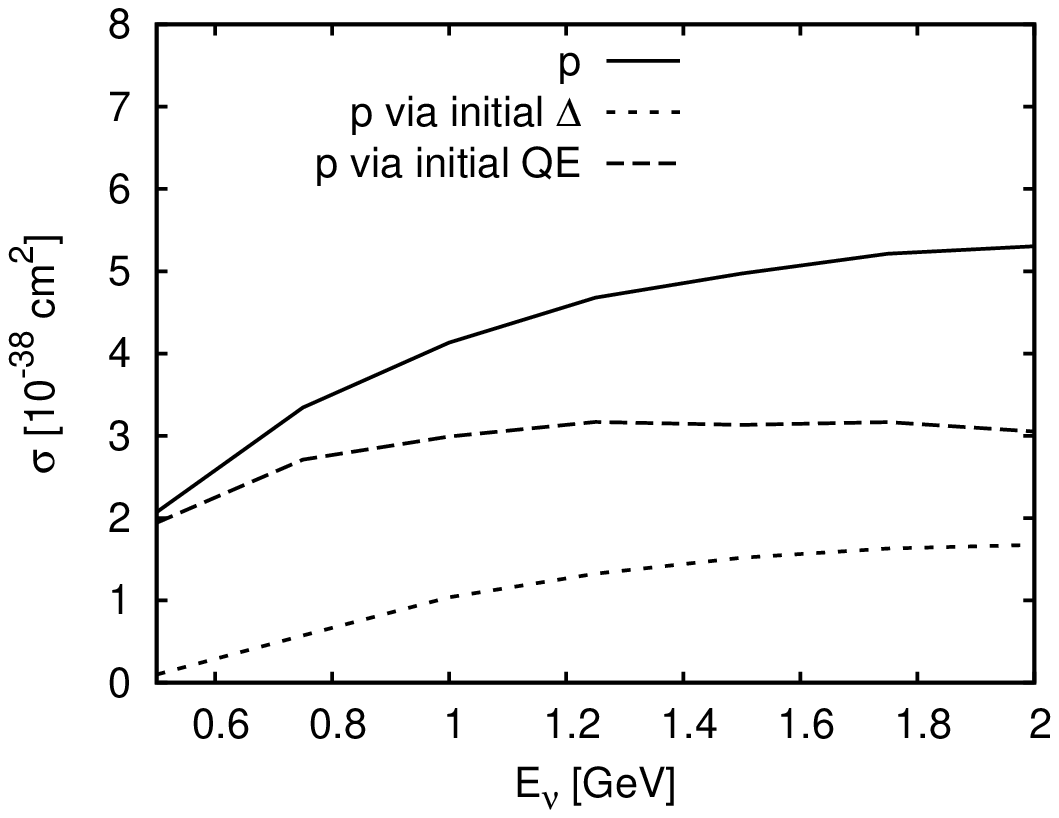}
 	\hspace{1em}
         \includegraphics[scale=\plotscale]{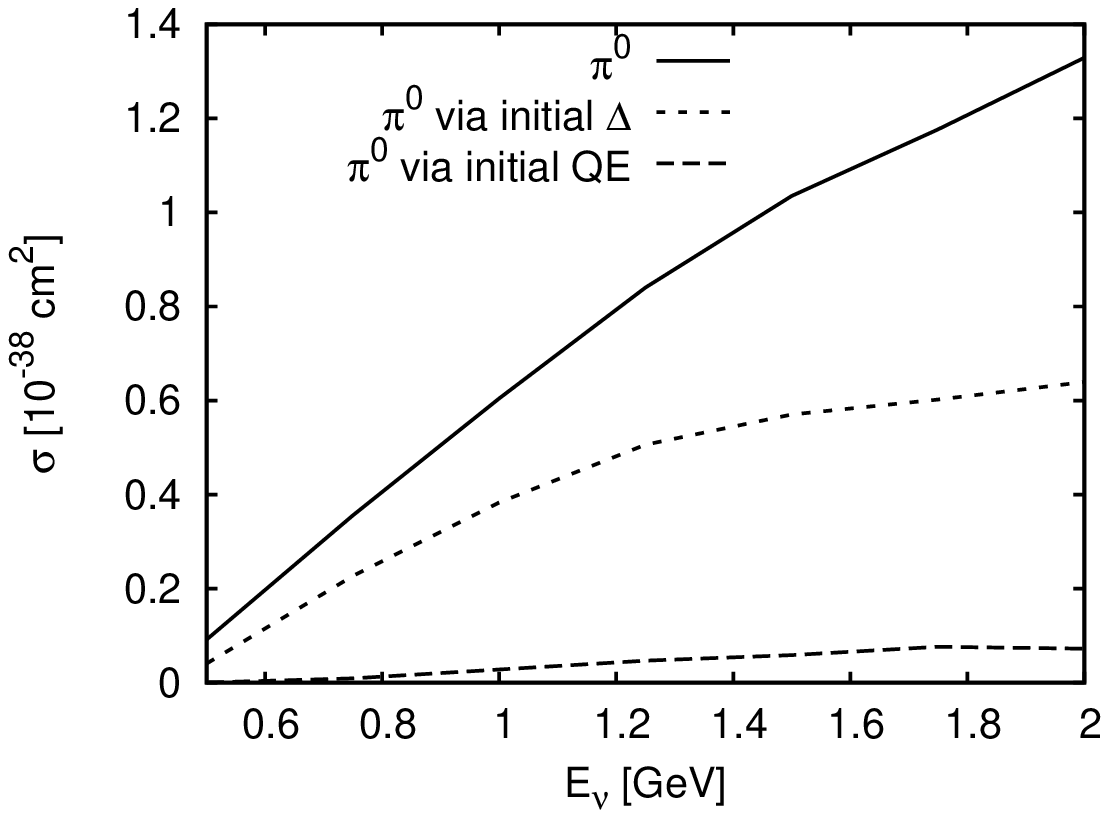}}
       \caption{Single proton knockout (left) and single $\pi^0$ production (right) cross sections for $\nu$ scattering off $^{12}$C (solid lines) through QE, resonance excitation and non-resonant processes. Possible origins, i.e.~the initial processes, are indicated (QE by long-dashed, $\Delta$ by short-dashed lines). Note the different scale.  \label{fig:FSIeffects}}
 \end{figure}

Within the GiBUU model we performed for the first time a systematic study of how quasielastic scattering and resonance excitation are interconnected by FSI, focusing, on one side, on resonance induced nucleon knockout and ``fake'' CCQE events and, on the other side, on side-feeding effects in pion production \cite{Leitner:2006ww,Leitner:2006sp}.
An example of the coupled channel effect in neutrino nucleus (here: $^{12}$C) reactions is given in \reffig{fig:FSIeffects} for $p$ and $\pi^0$ yields which are of interest in current LBL experiments. We find, that even though the proton knockout is dominated by protons coming from an initial QE reaction, the secondary protons from an initial $\Delta$ excitation contribute significantly.
FSI have also a considerable influence on CC neutrino induced pion production, where, due to isospin relations, $\pi^+$ dominate in the beginning. In the nucleus, they rescatter, get absorbed  and undergo charge exchange reactions leading to a disproportionate population of the $\pi^0$ channel. This has been earlier observed by Paschos \refetal{Paschos:2000be}. We stress that a correct understanding of the $\pi^0$ yield is required for a correct identification of the neutrino flavor in LBL experiments.

To conclude, we emphasize, that GiBUU is based on well-founded theoretical ingredients and it is the only model tested in various and very different reactions using the same physics input. In particular, an important prerequisite for any model for the interaction of neutrinos with nuclei is, that it provides a good description of electron- or photon-induced reactions. Within the GiBUU model, extensive tests against existing data are possible and have been successfully performed~\cite{Buss:2007ar,Buss:2008:Bormio,Alvarez-Ruso:2004ji}.
The GiBUU model is capable of incorporating the complexity of the nuclear many-body problem in an extensive open-source computer code which can be downloaded from our website~\cite{gibuu}.

\section{Applications: MiniBooNE and K2K}

We shall now present some examples for the application of our model.

\subsection{CCQE}

Charged current quasielastic (CCQE) events are commonly used in LBL experiments to determine the $\nu_\mu$ kinematics. Under the assumption that the nucleon is at rest within the nucleus, the neutrino energy has been reconstructed from QE events at the MiniBooNE experiment \cite{Aguilar:2007ru} using
\begin{equation}
  E_\nu = \frac{2(M_N - E_B)E_\mu - (E_B^2 - 2M_N E_B + m_\mu^2)} {2\:[(M_N - E_B) - E_\mu + \abskpr \cos\theta_\mu]},
\end{equation}
with a binding energy correction of $E_B=34\MeV$ and the measured muon properties. With that, we obtain the reconstructed $Q^2$ via
\begin{equation}
Q^2 = -m_\mu^2 + 2 E_\nu(E_\mu - \abskpr \cos\theta_\mu),
\end{equation}
Two immediate questions are raised by this procedure: (1) How good is the identification of CCQE events? (2) How exact is the crucial assumption of two body kinematics for nucleons bound in a nucleus where many in-medium modifications are present?

\begin{figure}
         \centerline{\includegraphics[scale=\plotscale]{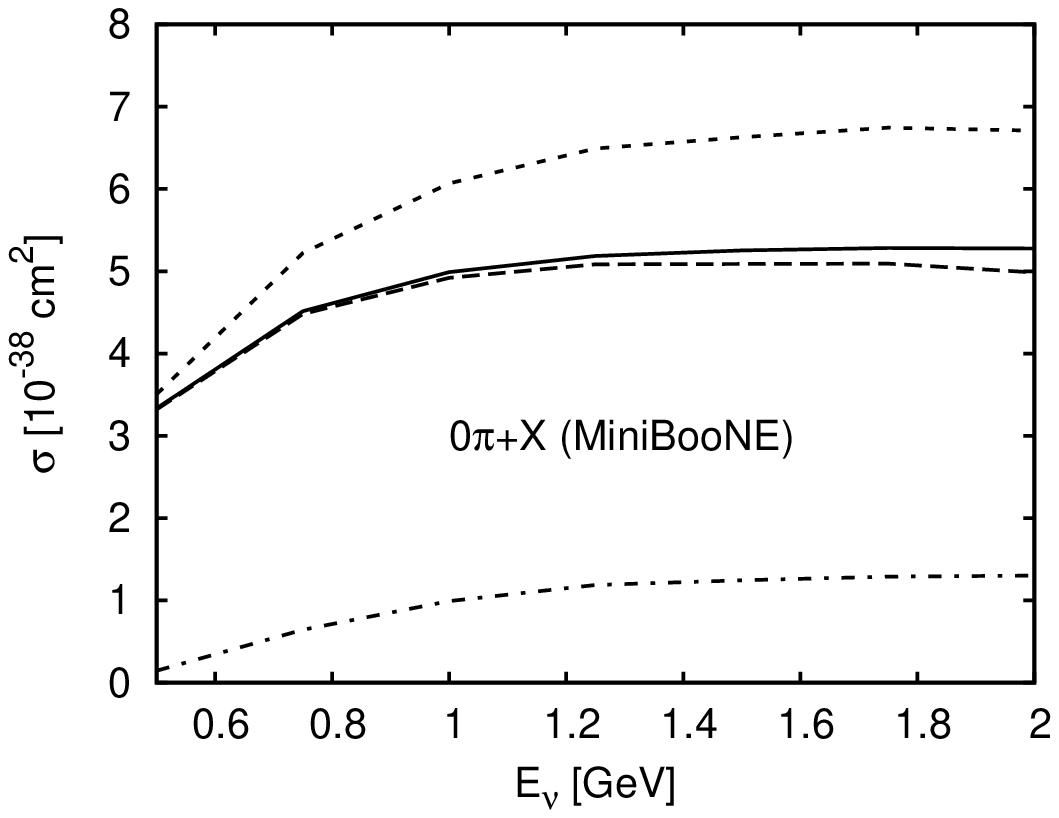}
 	\hspace{1em}
         \includegraphics[scale=\plotscale]{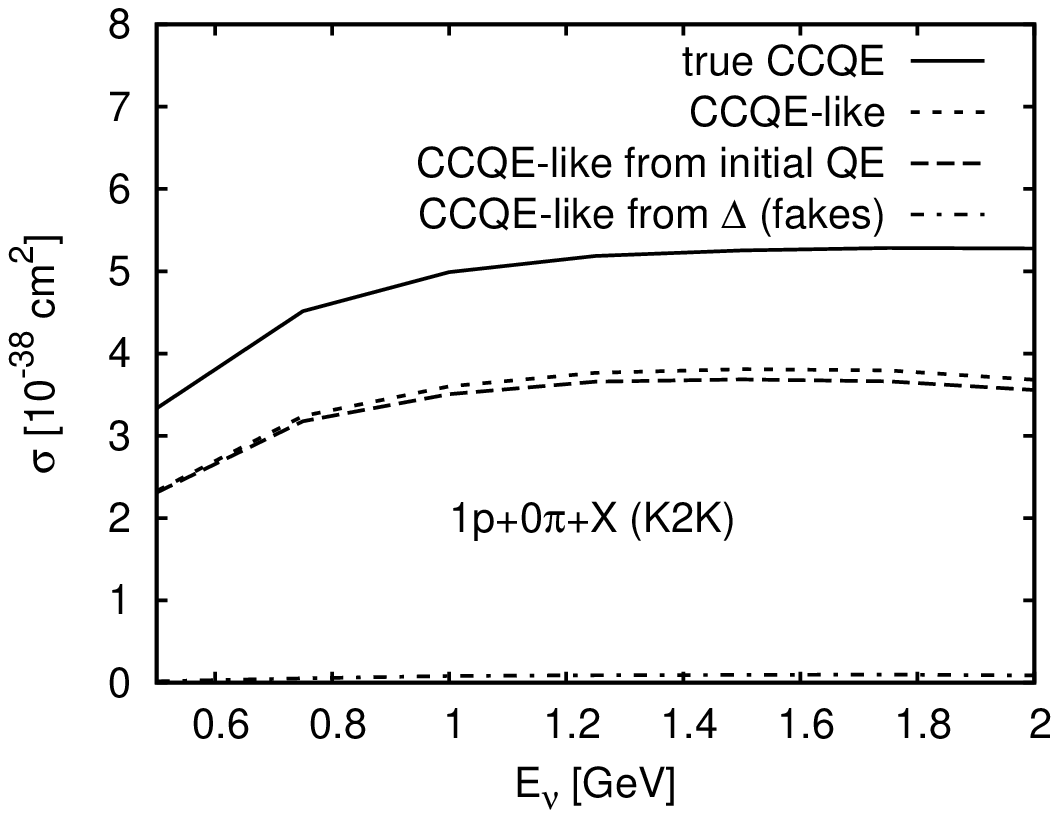}}
       \caption{Total QE cross section on $^{12}$C (solid line) compared to different methods on how to identify CCQE-like events in experiments (short-dashed lines). The contributions to the CCQE-like events are also classified (CCQE-like from initial QE (long-dashed), from initial $\Delta$ (dash-dotted)).   \label{fig:QEmethods}}
 \end{figure}

The experimental task is now to identify \emph{true} CCQE events in the detector, i.e., muons originating from an initial QE process. FSI might lead to misidentified events, e.g., an initial $\Delta$ whose decay pions are absorbed or which undergo ``pion-less decay'' can count then as CCQE event (we call this type of background events ``fake CCQE'' events). We denote every event which looks like a CCQE event by ``CCQE-like''. At MiniBooNE these are all the events where no pion is detected while at K2K these are all events where a single proton track is visible and at the same time no pions are detected. The two methods are compared in \reffig{fig:QEmethods}. The ``true'' CCQE events are denoted with the solid line, the CCQE-like events by the short-dashed one. Placing a cut only on pions, as MiniBooNE does, leads to a considerable amount of ``fake'' CCQE events (left panel, the short-dashed line is higher than the solid line). They are caused mainly by initial $\Delta$s via the mechanism described above; their contribution to the cross section is given by the dash-dotted line. On the contrary, less CCQE-like than true events are detected with the K2K method when one cuts both on pions and protons (left panel, difference between short-dashed and solid line).  The final state interactions of the initial proton lead to secondary protons, or, via charge exchange to neutrons which are then not detected as CCQE-like any more (\emph{single} proton track). We find that at K2K the amount of fake events in the CCQE-like sample is less than at MiniBooNE (compare difference between short-dashed and long-dashed line). We conclude that even if the additional cut on the proton helps to restrict the background, an error of about 25 \% remains, since the measured CCQE cross section underestimates the true one by that amount.

\begin{figure}
         \centerline{\includegraphics[scale=\plotscale]{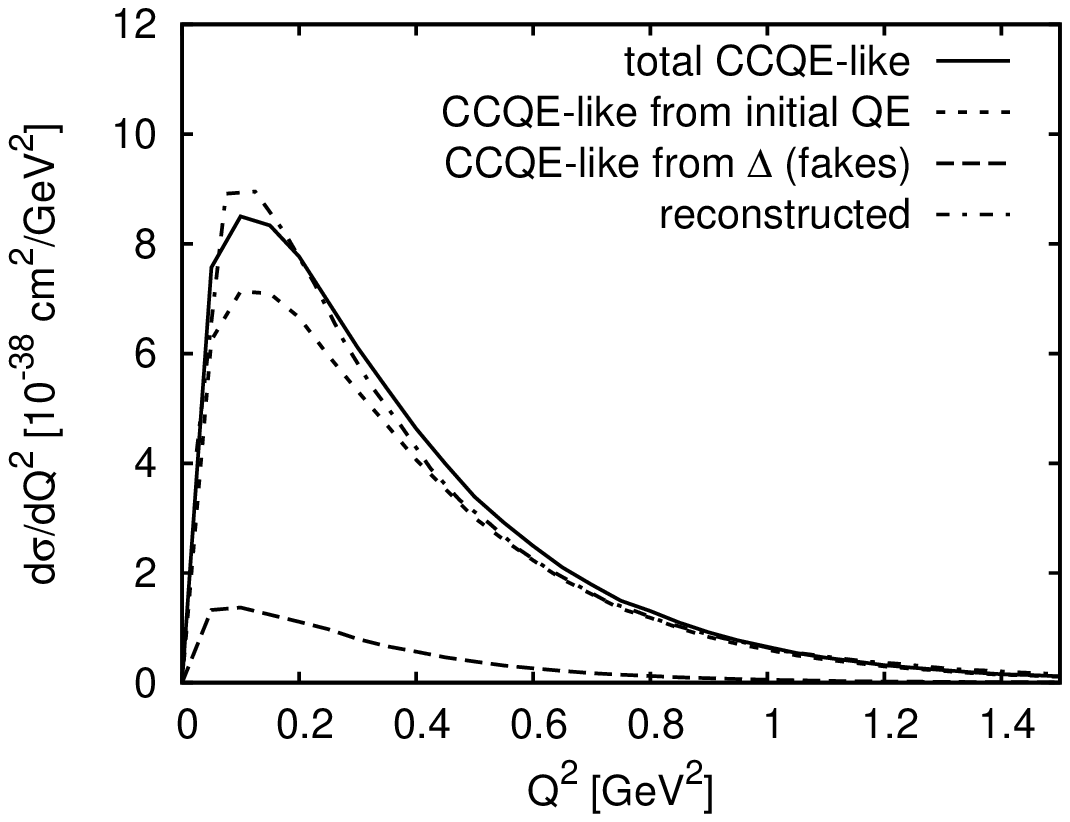}
           \hspace{1em}
         \includegraphics[scale=\plotscale]{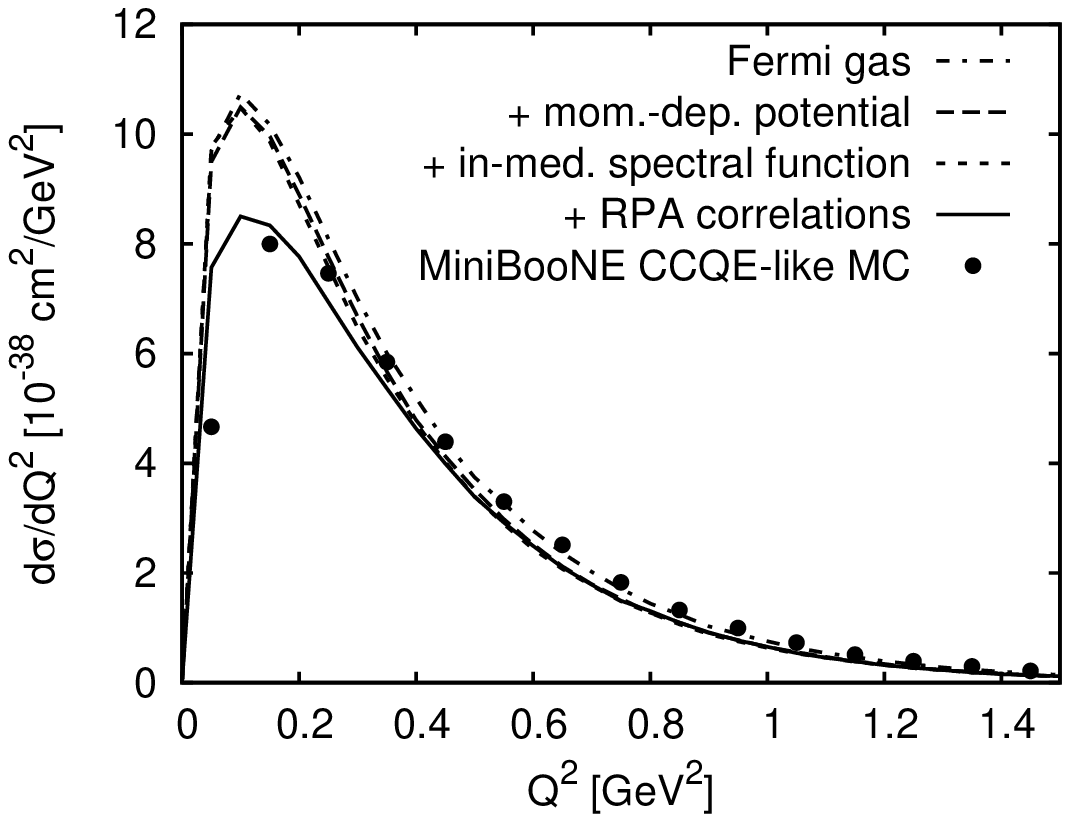}}
         \caption{Flux averaged $\dsdqs$  distribution of CCQE-like events at MiniBooNE. The left panel shows the composition of the cross section (initial QE, initial $\Delta$) and the reconstructed one. The right panel shows the influence of different in-medium modifications compared to the MiniBooNE tuned Monte Carlo output. For simplicity, only QE and $\Delta$ excitation were considered for the initial vertex. \label{fig:MiniBooNEqs}}
 \end{figure}

The flux averaged CCQE-like $Q^2$ distribution for MiniBooNE is shown in \reffig{fig:MiniBooNEqs}. The influence of the fake CCQE events on the energy reconstruction is can be inferred from the left panel. The distribution obtained by reconstructing $Q^2$ for the CCQE-like events via the formulas above is compared to the distribution of the true events (solid vs.~dash-dotted line). At lower $Q^2$ it is higher than the latter, but then it falls off faster. The difference between the two curves is caused by the fake CCQE events whose different muon kinematic affect the reconstruction. We also find that the reconstruction with the simplified formulas above turns out to be almost perfect when only true CCQE events (and not the whole CCQE-like sample) are taken into account.

In the right panel of \reffig{fig:MiniBooNEqs} we compare our calculation to the recent MiniBooNE findings: The MiniBooNE collaboration observed a dip at low $Q^2$ compared to their standard Monte Carlo prediction \cite{Aguilar:2007ru}. To reach agreement with the data, apart from changing the axial mass $M_A$ by about 25 \% with respect to values obtained in earlier experiments \cite{Kuzmin:2007kr}, they had to modify Pauli blocking in their Monte Carlo description! Since MiniBooNE has not provided absolute cross sections, a direct comparison to data is not yet possible. Instead, we compare to their CCQE-like Monte Carlo points\footnote{by normalizing the area} which includes the aforementioned tuning of parameters and treat this as experimental result. We compare these points to our calculations including different in-medium modifications on the cross section. While the momentum dependent potential and the spectral function had significant influence on the double differential distributions \cite{Buss:2007ar}, they are negligible here where we have integrated out one quantity, and do not improve the correspondence with the data. Polarization effects due to the strong interaction among nucleons modify the QE hadronic tensor. These are taken into account by including RPA correlations taken from Nieves \refetal{Nieves:2004wx} which lowers the spectrum at the peak and leads to a good description of the MiniBooNE points without tuning the axial mass or any other parameter. We shall explore this in more detail in a forthcoming publication.

 \subsection{CC1$\pi$}

LBL experiments demand for a realistic description of pion yields, for two main reasons: First, $\pi^0$s are an important background in $\nu_e$ appearance experiments which, through their photon decay, might affect the electron detection yielding also ``fake'' appearance events. Second, absorbed pions contribute to the CCQE-like background as discussed above.
Therefore, a good description of neutrino induced pion production in nuclei taking into account complex FSI is necessary and oversimplifications are not justified. We have shown \cite{Leitner:2007px} that, in particular, the ANP model applied by Paschos \refetal{Paschos:2000be} does not incorporate the well-known properties of the $N\Delta\pi$ dynamics in nuclei and, therefore, is not able to give reliable results for pion spectra in the energy region of interest.

\begin{figure}[tbp]
   \centerline{\includegraphics[scale=\plotscale]{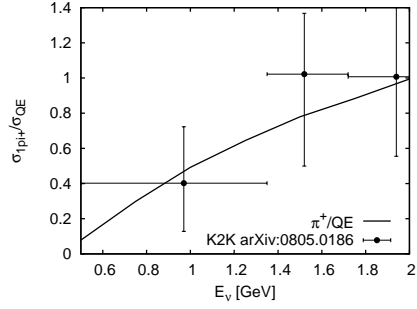}
   }
   \caption{Ratio of the total $\pi^+$ to the \emph{CCQE-like} yield in K2K compared to data taken from \refcite{Rodriguez:2008ea}. \label{fig:K2Kpion}}
 \end{figure}

As an example, in \reffig{fig:K2Kpion} we compare the recent K2K data for the ratio of $\pi^+$ to QE yields \cite{Rodriguez:2008ea} with the output of the GiBUU model and obtain good agreement.

\subsection{Radiative $\Delta$ decay}

MiniBooNE finds in its $\nu_\mu \to \nu_e$ oscillation result an excess of electron like events for neutrino energies less than $475\MeV$ which is not yet understood \cite{AguilarArevalo:2007it}. A possible source is the excitation of a $\Delta$ resonance via neutral current interaction followed by the radiative decay $\Delta \to \gamma N$. Since the MiniBooNE detector cannot distinguish between photons and electrons, this reactions gives rise to additional events in the low energy region. The major $\nu_\mu$-induced background, however, are $\pi^0$ coming from NC interactions detectable also via their photon decay products.
Of particular interest for experiments is thus how the photon to $\pi^0$ yield changes in the nuclear medium, depending on the $\Delta$ momentum and also the nuclear density.

In the vacuum, a rough estimate gives
\begin{equation}
\frac{\sigma_{\text{tot}}^\gamma}{\sigma_{\text{tot}}^{\pi^0}} (\Delta^{+/0})= \frac{0.0056}{(2/3)} = 0.008,
\end{equation}
where 0.0056 is the PDG branching fraction and 2/3 comes from the appropriate Clebsch-Gordon coefficient for $\Delta^{+/0} \to \pi^0 + N$. In the medium---we took $^{12}$C as used in MiniBooNE---our calculation has been performed as follows. First, $\Delta^0$ and $\Delta^+$ resonances are set inside the nucleus with momentum and radius chosen randomly within a given range. Then, they are propagated out taking into account all kind of decays and collisions. Afterwards, we calculate the total $\pi^0$ and the photon cross section as function of the \emph{initial} momentum (radius) of the $\Delta$. We take into account only those $\pi^0$ which actually made it out of the nucleus after the final state interactions. With that, we obtain for the above ratio $0.019$, which represents an increase of about a factor of 2.4 compared to the free case.

\begin{figure}
         \centerline{\includegraphics[scale=\plotscale]{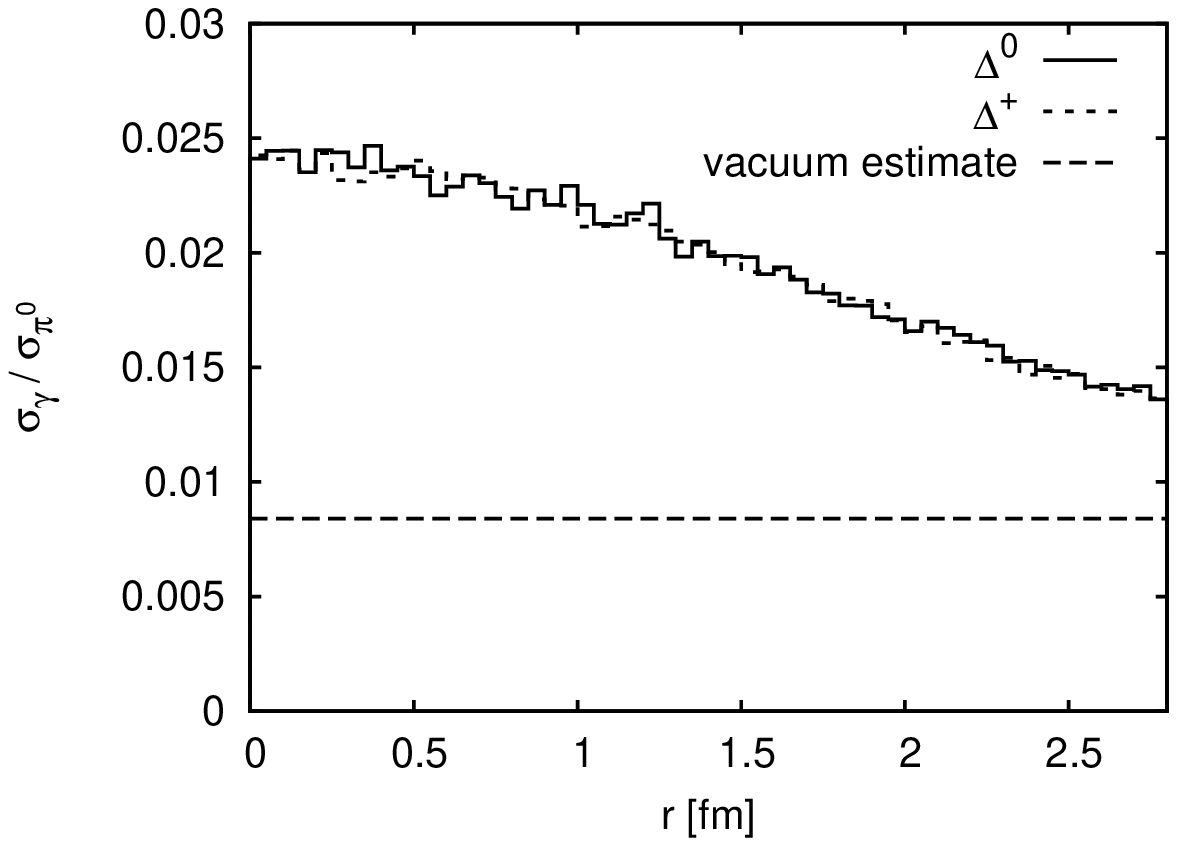}
 	\hspace{1em}
         \includegraphics[scale=.51]{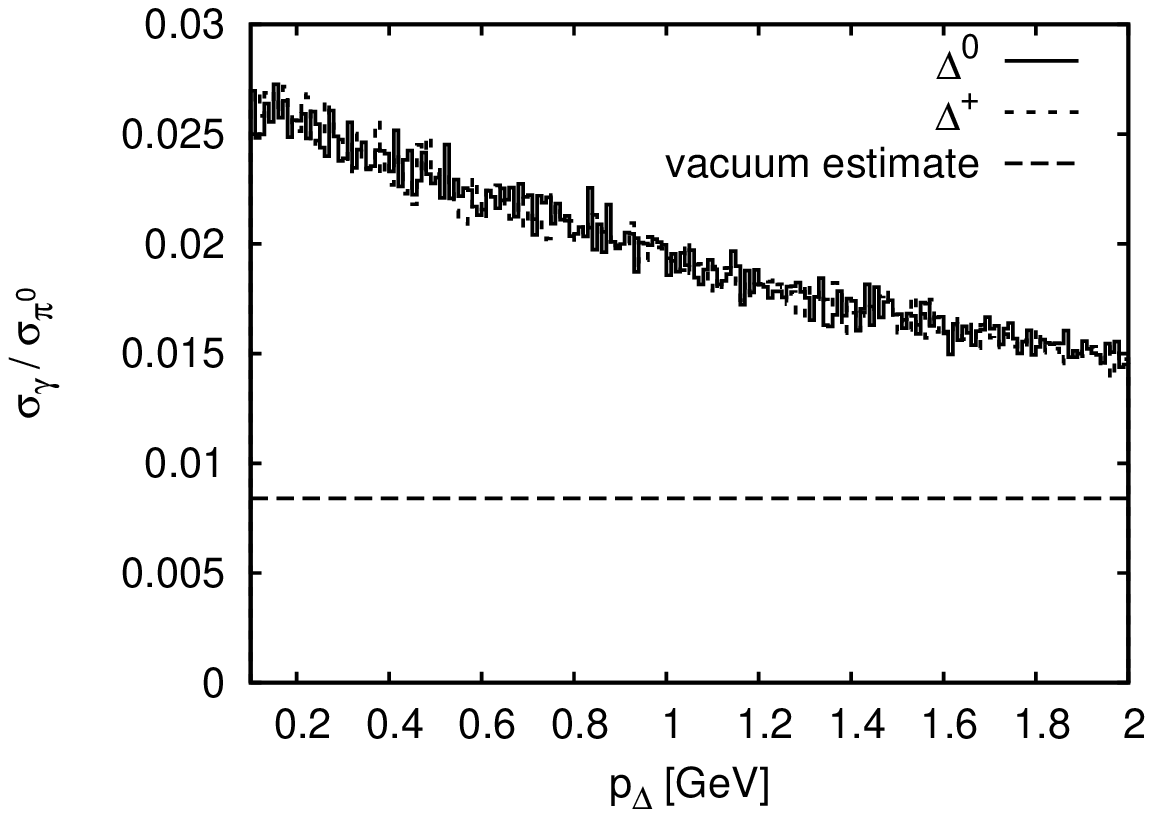}}
       \caption{Photon/$\pi^0$ rate changes in medium as a function of $\Delta$ position (left) and momentum (right).  \label{fig:radDelta}}
 \end{figure}

\reffig{fig:radDelta} shows how the photon/$\pi^0$ rate changes in medium as a function of $\Delta$ momentum and position (solid: initial $\Delta^0$, dashed: initial $\Delta^+$). In addition, the vacuum estimate is shown by the long dashed line.
In the momentum dependence one observes typical final state interaction effects: Slow $\Delta$s produce slow pions which are more likely to be absorbed in the medium than higher energetic ones which might pass through undisturbed. As expected, the medium modification is largest for those $\Delta$s which have been put in the middle of the carbon nucleus. However, one might expect, that the solid/dashed lines approach the vacuum value at a radius larger than the carbon radius. This does not happen here, because, as said before, we initialize the $\Delta$s at the beginning with a random momentum, therefore some of them can propagate into the nucleus and thus still undergo FSI which then again modify the spectrum.

To conclude, the production of photons vs.~$\pi^0$ is enhanced in the nuclear medium due to complex pion final state interactions reflecting in a strong dependence of density and momentum.

\section{Summary}

We conclude that in-medium effects in $\nu A$ scattering, and in particular FSI, are important for the interpretation of LBL
oscillation experiments. The influence of nuclear many-body effects and final state interactions have to be treated with the same degree of sophistication as the primary production process.

\paragraph{Acknowledgments.}
T.~L.~thanks T.~Katori from the MiniBooNE collaboration for fruitful discussions and for providing the Monte Carlo output. We thank G.~Garvey for stimulating communications on the radiative $\Delta$ decay. We thank all members of the GiBUU project for cooperation, in particular O.~Lalakulich for careful proofreading of this manuscript. This work was supported by the Deutsche Forschungsgemeinschaft.

%

\end{document}